\documentclass[12pt,twocolumn]{iopart}

\expandafter\let\csname equation*\endcsname=\relax
\expandafter\let\csname endequation*\endcsname=\relax
\usepackage{multirow}
\usepackage{array}
\usepackage{float}
\usepackage{lineno}
\usepackage{amsmath,amssymb,amsthm}
\usepackage{iopams}
\usepackage{epsfig}
\usepackage{booktabs}
\usepackage{subfigure}
\usepackage{graphicx}
\usepackage{dcolumn}
\usepackage{stmaryrd}
\usepackage{mathrsfs}
\usepackage{pifont}
\usepackage{bm}
\usepackage{latexsym}
\usepackage{hyperref}
\usepackage{newfloat,algcompatible}
\usepackage[size=small]{caption}
\usepackage{etoolbox}
\usepackage{newfloat}
\usepackage{color}
\usepackage{multirow}
\usepackage{varwidth}
\newtheorem{theorem}{Theorem}

\usepackage[T1]{fontenc}
\usepackage{epsfig,amsmath}
\makeatletter
\providecommand{\tabularnewline}{\\}
\newenvironment{cellvarwidth}[1][t]
{\begin{varwidth}[#1]{\linewidth}}
	{\@finalstrut\@arstrutbox\end{varwidth}}

\makeatother
\usepackage{babel}
\usepackage{lineno} 

\begin{document}
\title{Gradient-Based Excitation Filter for Molecular Ground-State Simulation}

\author{Runhong He$^1$, Qiaozhen Chai$^1$, Xin Hong$^1$, Ji Guan$^1$, Guolong Cui$^2$, Shengbin Wang$^3$, Shenggang Ying$^{1*}$}

\address{
	1. Key Laboratory of System Software (Chinese Academy of Sciences) and State Key Laboratory  of Computer Science, Institute of Software, Chinese Academy of Sciences,	Beijing 100190, China \\
	2. Arclight Quantum Co., LTD. Chinese Academy of Sciences, Beijing	101408, China\\
	3. China Telecom Quantum Information Technology Group Co., LTD, Hefei 233000, China}

\ead{$^*$yingsg@ios.ac.cn}

\vspace{10pt}
\begin{indented}
	\item[]July  2025
\end{indented}
\begin{abstract}
Molecular ground-state simulation is one of the most promising fields for demonstrating practical quantum advantage on near-term quantum computers. However, the Variational Quantum Eigensolver (VQE), a leading algorithm for this task, still faces significant challenges due to excessive circuit depth. This paper introduces a method to
efficiently simplify the Unitary Coupled-Cluster with Single and Double
Excitations (UCCSD) ansatz on classical computers. We propose to estimate
the correlation energy contributions of excitations using
their gradients at Hartree-Fock state, supported by a theoretical
proof. 
For molecular systems with $K$ orbitals, these gradients can be obtained with complexity only $O(K^8)$, which can be efficiently implemented  on classical computers, especially in parallel.
By sorting and truncating the excitations based on these gradients, the simplified ansatz can be obtained immediately, avoiding the challenging task of optimizing ansatz structure on a quantum computer.
Furthermore, we introduce a strategy to indirectly identify critical excitations through spin-adapted constraints, reducing gradient computations by $60\%$.
Numerical experiments on prototype molecular systems (H${_4}$, HF, H${_2}$O, BeH${_2}$
and NH$_3$) demonstrate that our approach achieves up to $46\%$ parameter decrease, $60\%$ circuit depth reduction and $678\times$ runtime speedup compared to the state-of-the-art ADAPT-VQE algorithm, enabling significantly more compact quantum circuits with enhanced near-term feasibility. 
\end{abstract}
\vspace{2pc}

\section{INTRODUCTION}

The chemical properties of matter are  fundamentally determined by molecular
energy levels \cite{Quantum_computational_chemistry}. This relationship is particularly evident in chemical kinetics, where accurate prediction of reaction rates requires precise knowledge of ground-state energies \cite{Eyring_equation}. Within classical computational frameworks, molecular energies can be exactly determined through the Full Configuration Interaction (FCI) method \cite{review_molecular_calculations}. However, the FCI method's computational cost scales exponentially with system size, making it feasible only for small molecular systems \cite{szabo}. For larger molecular systems, approximations have to be introduced to balance computational cost and accuracy. A prominent example is the Coupled-Cluster Singles and Doubles (CCSD) method, which truncates the cluster operator at single and double excitations, significantly reducing computational complexity. As low energy excitations dominate the ground state in many systems, this truncation can still yield good approximations to the ground-state energy \cite{Quantum_computational_chemistry, szabo}.

Quantum algorithms that leverage the properties of quantum superposition
and entanglement can achieve significant speedup over classical algorithms
in solving certain important problems \cite{qc_nielsen}.
The first quantum algorithm capable of solving Hamiltonian eigenvalue
problems is the Quantum Phase Estimation (QPE) algorithm \cite{QPE}.
Compared to classical algorithms, QPE can provide exponential speedup.
However, the quantum hardware requirements for QPE -- including error
correction mechanisms and long coherence times -- currently exceed
the capabilities of near-term Noisy Intermediate-Scale Quantum (NISQ) devices \cite{NISQ}. Recently, the Variational
Quantum Eigensolver (VQE) algorithm \cite{VQE_Science,VQE_photon,VQE_big_size_Nature_Comm_2022,VQE},
which adopts a quantum-classical hybrid architecture, has attracted
wide-spread attention. In VQE, the second-quantized Hamiltonian can
be transformed into Pauli strings through established encodings such
as Jordan-Wigner \cite{Jordan-Wigner} or Bravyi-Kitaev \cite{Bravyi-Kitaev},
thereby enabling the expectation value of the trial state with respect
to the Hamiltonian to serve as the eigenvalue. During practical execution
of VQE, the quantum processor runs a parameterized quantum circuit
(ansatz circuit) to generate trial states, while the classical computer
calculates the loss function and gradients based on measurement outcomes.
These results are then used to update the parameters in the ansatz
circuit to obtain improved solutions. Compared to the QPE algorithm,
VQE features shallower circuit depths and holds promise for practical implementation
in the NISQ era.

The design of the ansatz circuit lies at the heart of the VQE.
In contrast to the hardware-efficient ansatz \cite{HEA_nature} --
which, while readily implementable, is plagued by the barren plateau
problem \cite{Barren_plateaus} -- the Unitary Coupled-Cluster with
Single and Double Excitations (UCCSD) method \cite{Quantum_computational_chemistry}
has attracted extensive attention and in-depth investigation, owing
to its clear physical interpretation and robust convergence properties.
The UCCSD ansatz requires transformation of fermionic excitations
into parameterized quantum circuits through two primary approaches.
(1) Conventional Pauli-based implementation \cite{qubit-excitation-operator,qubit_ADAPT_VQE}:
excitations are first mapped to Pauli strings via encodings \cite{Jordan-Wigner,Bravyi-Kitaev},
which are then synthesized into quantum circuits through exponentiation
techniques \cite{qc_nielsen}. (2) Emerging direct construction methods
\cite{efficient_excitations_circ,exciation_9_cnot,adapt_9_cnot}:
these circumvent the Pauli decomposition by directly implementing
excitations via combining $C^nR_{y}$ gate and basis transformation
operations, thereby significantly improving computational efficiency.

The implementation cost of the UCCSD ansatz is dominated
by double excitations, whose number scales with system
size as $O(K^{4})$ for $K$ molecular orbitals. This scaling inevitably
generates quantum circuit depths surpassing the error tolerance limits
of NISQ hardware, precluding simulations of industrially meaningful
molecules. It is crucial to emphasize that not all excitations in
the UCCSD ansatz contribute equally -- a substantial fraction have
only negligible effects on ground-state energy. Therefore, accurately
identifying and filtering out redundant excitations in the UCCSD ansatz
is essential for reducing quantum circuit depth and advancing the
practical application of VQE algorithms. 

To address this problem, Ref.~\cite{VQE_point_group_symmetry} proposed
a compression method named ``SymUCCSD'' based on molecular point
group symmetry. SymUCCSD systematically eliminates symmetry-violating
excitations, but suffers from two limitations: (1) It's efficacy strongly
depends on molecular symmetry -- it achieves substantial ansatz compression
for highly symmetric systems but shows limited improvement for low-symmetry
molecules. (2) It retains symmetry-compliant yet energetically negligible
(``non-important'') excitation operators,
leaving residual redundancies in the simplified ansatz.

The ADAPT-VQE algorithm \cite{ADAPT-VQE} provides an innovative
approach to excitation selection. It utilizes an ``Excitation Pool''
containing all UCCSD excitations and constructs a compact
quantum circuit from identity by dynamically adding only the most
energetically significant excitations (selected based on gradient magnitudes)
at each iteration. Compared to fixed ansatz such as the standard UCCSD
and SymUCCSD, ADAPT-VQE achieves chemical accuracy with significantly
fewer parameters while maintaining comparable precision. Furthermore,
this method demonstrates remarkable generality, being applicable to
ground-state energy calculations across diverse molecular systems.
These advantages have propelled the ADAPT-VQE to rapid widespread
adoption since its inception, establishing it as a leading approach in the field \cite{qubit_ADAPT_VQE, adapt_9_cnot, adapt_symmetry_breaking, vqe_review, vqe_review2, adapt_vqe_application, adapt_pool_size}. 
However, the ADAPT-VQE
suffer from non-negligible inefficiencies that fundamentally limit
their applicability to larger molecular systems: each iteration requires
(1) evaluating the gradients of all excitations in the Excitation
Pool in current trial state, and (2) re-optimizing all parameters
in the expanded ansatz after adding a new excitation. Specifically,
for an UCCSD ansatz containing $N$ excitations (of which $M$ are
energetically significant), the ADAPT-VQE algorithm requires $O(M)$
VQE optimization procedures, accumulatively optimizes $O(M^{2})$
parameters in total, and performs $O(N\times M)$ excitation gradient
evaluations during the selection process. Here, a ``VQE optimization procedure'' refers to the end-to-end computational process encompassing:  (1) ansatz parameter initialization, (2) iterative parameter updates via hybrid quantum-classical feedback, and (3) convergence verification against predefined thresholds. 

In this work, we establish a theoretical connection between the excitations'
gradients at Hartree-Fock state and their correlation energy contributions,
thereby developing an efficient excitation selection algorithm. We term this algorithm GBEF (Gradient-Based Excitation
Filter). After compressing both the system Hamiltonian and Hartree-Fock state into the excitation's subspace, each gradient can be computed with complexity comparable to $16\times16$ matrix
multiplication. 
This contrasts with ADAPT-VQE, which approximates gradients for quantum hardware-prepared ansatz states through costly measurements. 
GBEF leverages these gradients to systematically sort and truncate redundant excitations, achieving significant efficiency gains over iterative ADAPT-VQE. 
For a $K$-orbital molecular system, GBEF achieves $O(K^8)$ computational complexity, allowing efficient parallel implementation on classical computers.
Moreover, our method requires no prior knowledge of molecular symmetry,
making it applicable to a wider range of molecular systems than SymUCCSD. Finally, we introduce a parameter-based
strategy for indirectly selecting critical excitations, leveraging spin-adaption constraints to reduce the number of gradient calculations by about $60\%$.

\section{Methods}

\subsection{Specification of the adopted notations}
To construct the method, we introduce the following definitions and
notation. First, the occupied molecular orbitals are indexed with
subscripts $a$ and $b$, while virtual orbitals are indexed with
$r$ and $s$.

In UCCSD, the trail state $|\Psi(\vec{\theta})\rangle$ is prepared
from the Hatree-Fock state $|HF\rangle$ by an unitary evolution operator
(ansatz) $\text{e}^{T(\vec{\theta})}$ involving single and double
excitations \cite{Quantum_computational_chemistry}:
\begin{equation}
	|\Psi(\vec{\theta})\rangle=\text{e}^{T(\vec{\theta})}|HF\rangle,
\end{equation}
with the exponent 
\begin{equation}
	T(\vec{\theta})=\sum_{r,a}\theta_{a}^{r}A_{a}^{r}+\sum_{r>s,a>b}\theta_{ab}^{rs}A_{ab}^{rs}.\label{eq:ansatz}
\end{equation}
The single and double excitations are defined as
\begin{equation}
	A_{a}^{r}=\hat{a}_{r}^{\dagger}\hat{a}_{a}-\hat{a}_{a}^{\dagger}\hat{a}_{r},\quad A_{ab}^{rs}=\hat{a}_{r}^{\dagger}\hat{a}_{s}^{\dagger}\hat{a}_{a}\hat{a}_{b}-\hat{a}_{a}^{\dagger}\hat{a}_{b}^{\dagger}\hat{a}_{r}\hat{a}_{s},
\end{equation}
where operators $\hat{a}_{i}^{\dag}$ (creation) and $\hat{a}_{i}$
(annihilation) obey fermionic anti-commutation rules:
\begin{equation}
	\{\hat{a}_{i},\hat{a}_{j}^{\dag}\}=\delta_{i,j},\quad\{\hat{a}_{i},\hat{a}_{j}\}=\{\hat{a}_{i}^{\dag},\hat{a}_{j}^{\dag}\}=0.
\end{equation}

Under first-order Trotter-Suzuki approximation \cite{trotter}, the
evolution operator can be decomposed as:
\begin{equation}
	\text{e}^{T(\vec{\theta})}\thickapprox\prod_{k}\text{e}^{\theta_{k}A_{k}},
\end{equation}
where each $A_{k}$ represents either a single excitation $A_{a}^{r}$,
or a double excitation $A_{ab}^{rs}$. The quantum circuit implementation
of excitations has been extensively investigated in the literature,
featuring various overheads that include: encodings \cite{Jordan-Wigner,Parity_basis,Bravyi-Kitaev},
qubit excitations \cite{qubit-excitation-operator} and efficient
circuit designs \cite{efficient_excitations_circ,adapt_9_cnot,exciation_9_cnot}.
This work focuses only on simplifying the UCCSD ansatz by reducing
the number of excitations, without considering their implementation
details.

In conventional VQE implementations, excitation parameters are typically
optimized as independent variables. This approach, while flexible,
does not inherently preserve the spin symmetry of the resulting wavefunction,
as the final trial state may not remain an eigenstate of the $\hat{S^{2}}$
operator. For systems where the exact ground state is expected to
maintain definite spin symmetry (particularly in closed-shell configurations),
we can impose the spin-adaptation constraints on the excitations.
These constraints enforce commutativity between the excitations and
$\hat{S^{2}}$, thereby guaranteeing spin-preserving solutions while
simultaneously reducing the number of free variational parameters. By reinterpreting
the subscripts in Eq.~(\ref{eq:ansatz}) to denote spatial (rather than
spin) orbitals and explicitly indicating spin flavors ($\alpha/\beta$),
we can derive explicit spin-adaptation constraints for closed-shell
configurations \cite{spin_adaptation}:
\begin{equation}
	\theta_{a_\alpha}^{r_\alpha}=\theta_{a_\beta}^{r_\beta},
\end{equation}
\begin{equation}
	\theta_{a_\alpha b_\alpha}^{r_\alpha s_\alpha}=\theta_{a_\beta b_\beta}^{r_\beta s_\beta}=\theta_{a_\alpha b_\beta}^{r_\alpha s_\beta}+\theta_{a_\alpha b_\beta}^{r_\beta s_\alpha},
\end{equation}
\begin{equation}
	\theta_{a_\alpha b_\beta}^{r_\alpha s_\beta}=\theta_{a_\beta b_\alpha}^{r_\beta s_\alpha}\quad\&\quad\theta_{a_\alpha b_\beta}^{r_\beta s_\alpha}=\theta_{a_\beta b_\alpha}^{r_\alpha s_\beta},
\end{equation}
\begin{equation}
	\theta_{a_\alpha b_\beta}^{r_\alpha r_\beta}+\theta_{a_\beta b_\alpha}^{r_\alpha r_\beta}=\theta_{a_\alpha a_\beta}^{r_\alpha s_\beta}+\theta_{a_\alpha a_\beta}^{r_\beta s_\alpha}=0.
\end{equation}
While the spin-adaptation constraints do not reduce the number of
excitations in the ansatz, they significantly decrease the number
of free parameters to be optimized, thereby reducing the measurement cost
in VQE optimization. Compared to conventional approaches treating
each excitation individually, systematic application of spin-adaptation
constraints typically enables a $\sim60\%$ reduction in parameter
counts while preserving wavefunction accuracy.

In the framework of VQE, the ground-state engergy is obtained by minizing
the expectation value of the Hamitonian with respect to the ansatz
parameters:
\begin{equation}
	E=\min_{\vec{\theta}}\langle\Psi(\vec{\theta})|\mathscr{H}|\Psi(\vec{\theta})\rangle.
\end{equation}
Within the Born-Oppenheimer approximation, the second-quantized electronic
Hamiltonian of a molecule in atomic units takes the form \cite{Quantum_computational_chemistry}
\begin{equation}
	\mathscr{H}=\sum_{a,r}h_a^r\hat{a}_{r}^{\dagger}\hat{a}_{a}+\frac{1}{2}\sum_{a,b,r,s}h_{ab}^{rs}\hat{a}_{r}^{\dagger}\hat{a}_{s}^{\dagger}\hat{a}_{a}\hat{a}_{b},\label{eq:ham}
\end{equation}
with one- and two-electron integrals in a spin-orbital basis \cite{quantum_chemetry_book_2}
\begin{equation}
	h_a^r=\int d\mathbf{x}\phi_{r}^{*}(\mathbf{x})\left(-\frac{\nabla^{2}}{2}-\sum_{I}\frac{Z_{I}}{\left|\mathbf{r-R}_{I}\right|}\right)\phi_{a}(\mathbf{x}),
\end{equation}
\begin{equation}
	h_{ab}^{rs}=\int \text{d}\mathbf{x}_{1}\text{d}\mathbf{x}_{2}\frac{\phi_{r}^{*}(\mathbf{x}_{1})\phi_{s}^{*}(\mathbf{x}_{2})\phi_{a}^{*}(\mathbf{x}_{2})\phi_{b}^{*}(\mathbf{x}_{1})}{\left|\mathbf{r}_{1}-\mathbf{r}_{2}\right|}.
\end{equation}
where $Z_{I}$, $\mathbf{R}_{I}$ and $\mathbf{r}_{1}$ denote the
atomic number and position of the $I$-th nucleus and position of
the $i$-th electron, respectively. These integrals can be efficiently
computed within specified basis sets, such as the STO-3G basis set.
When calculating the expectation value of the Hamiltonian (\ref{eq:ham}),
it is necessary to transform it from the second-quantized form to
a Pauli string representation through encodings \cite{Jordan-Wigner}.

\subsection{Gradient-Based Excitation Filter Algorithm}
To simplify the UCCSD ansatz, we establish the following theorem
(proof is provided in Appendix~\ref{sec:Appendix}). 

\begin{theorem} \label{theorem1} If the Hartree-Fock state $|HF\rangle$ is a good approximation to the exact ground state wave-function $|\Phi_{0}\rangle$, an excitation's contribution to the correlation energy is related to the gradient of the system energy involved only this excitation at the $|HF\rangle$ with respect to the coefficient $\theta=0$.
\end{theorem}

This theorem points to an efficient path to filter out redundant excitations
in the UCCSD ansatz. Since double (single) excitations act on only
4 (2) qubits, we can easily compute excitations'
gradients by compressing the full system Hamiltonian and Hartree-Fock state
into this small subspace. This enables efficient estimation of each
excitation's relative contribution to the correction energy through
the Thm.~\ref{theorem1}. 

\begin{figure*}
	\centering
	\includegraphics[scale=0.12]{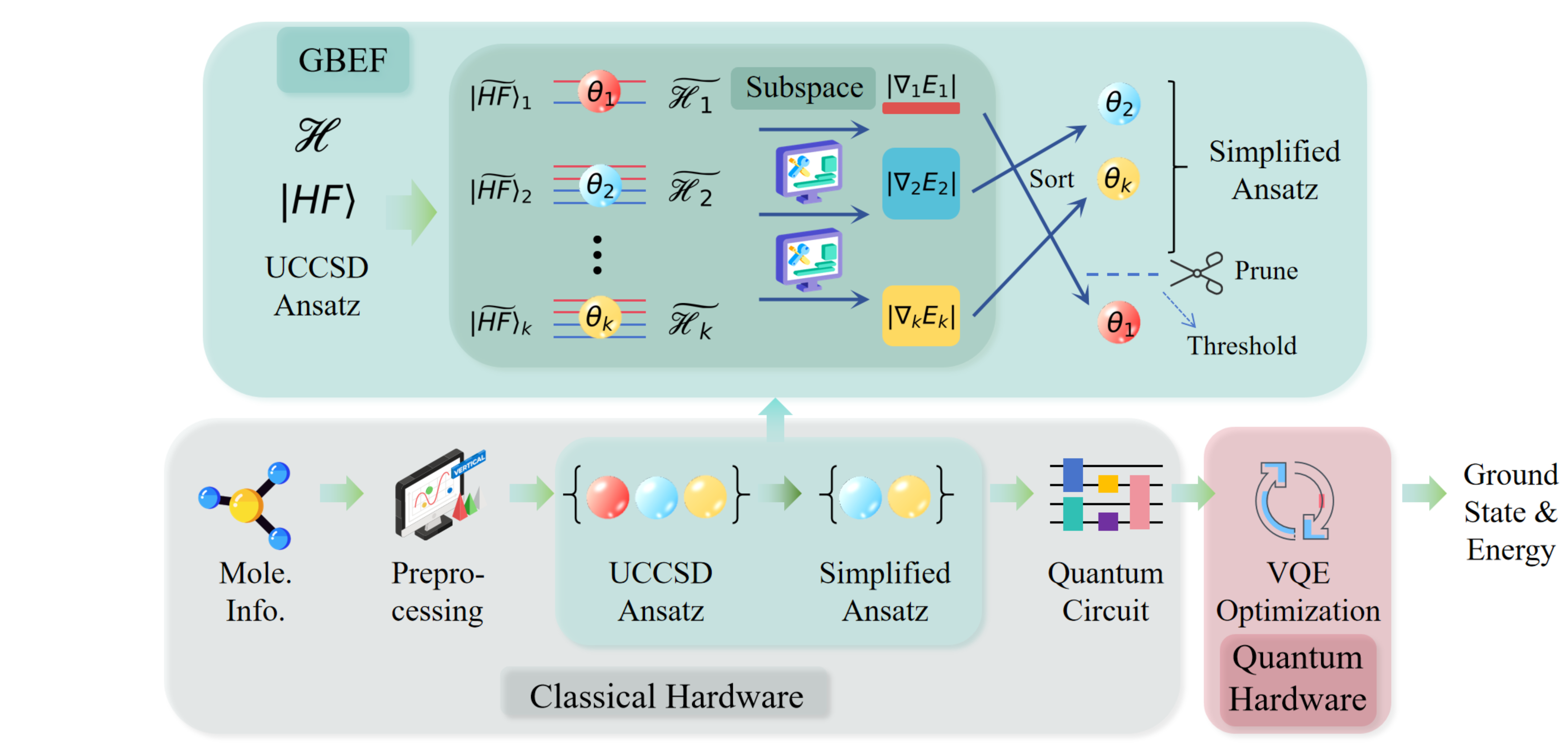}
	\caption{Schematic depiction of the GBEF algorithm and its integration with
		the standard VQE workflow.}\label{fig:workflow}
\end{figure*}

The overview of our GBEF algorithm and its integration with the standard
VQE workflow is drawn schematically in Fig.~\ref{fig:workflow}
and is as follows:

(1) According to the molecular information, compute one- and two-electron
integrals in Eq.~(\ref{eq:ham}) at a chosen basis set, and apply fermion-to-qubit encoding
(e.g., the Jordan-Wigner encoding \cite{Jordan-Wigner}) to obtain
the full Hamiltonian in Pauli representation. Construct the Hatree-Fock state $|HF\rangle$ and the UCCSD ansatz.

(2) Randomly assign to each parameter $\theta_{k}$ in the UCCSD ansatz
an excitation $A_{k}$ that is only associated with $\theta_{k}$.
These parameter-excitation pairs collectively form a set denoted as
$\{(\theta_{k},A_{k})\}$.

(3) For each element in $\{(\theta_{k},A_{k})\}$, project: the excitation
itself ($A_{k}\rightarrow\widetilde{A_{k}}$), the full Hamitonian
($\mathscr{H}\rightarrow\widetilde{\mathscr{H}_{k}}$), the Hartree-Fock state
($|HF\rangle\rightarrow|\widetilde{HF}\rangle_{k}$) onto the corresponding
4-qubit (2-qubit) subspace for double (single) excitation, thereby forming a new set denoted as $\{(\theta_{k},\widetilde{A_{k}},\widetilde{\mathscr{H}_{k}},|\widetilde{HF}\rangle_{k})\}$.

(4) For each element in $\{(\theta_{k},\widetilde{A_{k}},\widetilde{\mathscr{H}_{k}},|\widetilde{HF}\rangle_{k})\}$,
compute system energy's gradient with respect to $\theta_{k}=0$ at $|\widetilde{HF}\rangle_{k}$
via
\begin{equation}
	\nabla_{k}E_{k}:=\frac{\text{d}E_{k}}{\text{d}\theta_{k}}\Big|_{\theta_{k}=0}=\hspace{0cm}_{k}\langle\widetilde{HF}|[\widetilde{\mathscr{H}_{k}},\widetilde{A_{k}}]|\widetilde{HF}\rangle_{k},
\end{equation}
where $E_k:=\hspace{0cm}_{k}\langle\widetilde{HF}|\text{e}^{-\theta_k \widetilde{A_{k}}}\widetilde{\mathscr{H}_{k}}\text{e}^{\theta_k \widetilde{A_{k}}}|\widetilde{HF}\rangle_{k}$.
This step can be easily calculated  with
computational complexity comparable to $16\times16$ ($4\times4$) matrix
multiplication for double (single) excitations.

(5) List and sort all parameters $\theta_{k}$ in descending order
based on their corresponding absolute gradient values $\left|\nabla_{k}E_{k}\right|$.

(6) Prune the parameter sequence according to a certain strategy, yielding an optimized parameter
subset $\{\theta_{k}\}_{\text{opt}}$.

(7) Collect all excitations in UCCSD that associated with parameters
in $\{\theta_{k}\}_{\text{opt}}$, and combine them to form the simplified
ansatz, thereby generating the corresponding quantum circuit.

(8) Perform standard VQE optimization procedure using the constructed
quantum circuit to determine the molecular ground-state energy.

As described above and illustrated in Fig.~\ref{fig:workflow},
the distinctive workflow of our GBEF algorithm for UCCSD ansatz simplification
includes operates from Step 2 to Step 7. Specifically, in Steps
2 and 7, we propose an excitation-grouping strategy based on free
parameters that fully exploits spin-adaptation constraints to reduce
gradient computations by about $60\%$. In Steps 3-4, we compress
both the full Hamiltonian and Hartree-Fock state into the subspace of excitations, enabling efficient classical computation of parameter gradients. Since the gradient computation for each excitation incurs fixed costs, the dominant computational expense in GBEF stems from the Hamiltonian compression. For a system with $K$ molecular orbitals, where both the number of excitations and terms in Hamiltonian scale as $O(K^4)$, the overall complexity of these two steps is $O(K^8)$ –- a process can be efficiently handled by classical computers, especially in parallel.
In Steps 5-6, we sort and truncate parameters according to the absolute
values of their gradients, thereby identifying the unessential parameters
to filter out. The truncation threshold can be adjusted to balance
accuracy and efficiency. Steps 1 and 8 ensures end-to-end compatibility
with the existing VQE workflow.

\section{Results}
In this section, we investigate the performance of our GBEF algorithm
in finding equilibrium geometries for 5 typical molecules: H$_4$,
HF, H$_2$O, BeH$_2$, and NH$_3$.
Tab.~\ref{tab:mole} comprehensively characterizes the studied molecular systems, including bond types, equilibrium bond lengths and scanned bond length ranges for
geometry optimization.
The number of qubits required for VQE implementation serves as a 
metric for molecular system size.

\begin{table}[t]
	\centering
	\caption{ Qubit requirements, bond types and bond length ranges for quantum simulations of the investigated molecules. 21 equally spaced points are sampled across the range of bond distances to draw a statistical conclusion.}\label{tab:mole}
	\begin{tabular}{ccccc}
		\hline 
		Molecule & \begin{cellvarwidth}[t]
			\centering
			Qubit Count\\in VQE
		\end{cellvarwidth} & \begin{cellvarwidth}[t]
			\centering
			Bond\\Type
		\end{cellvarwidth} & \begin{cellvarwidth}[t]
			\centering
			Equilibrium Bond\\Length ($\mathring{\text{A}})$
		\end{cellvarwidth} & \begin{cellvarwidth}[t]
			\centering
			Bond Length\\ Range ($\mathring{\text{A}})$
		\end{cellvarwidth}\tabularnewline
		\hline 
		$\mathrm{{H_{4}}}$ & 8 & H-H & 0.89 & $[0.1,1.1]$ \\
		$\mathrm{{HF}}$ & 12 & F-H & 0.99 & $[0.1,1.1]$ \\
		$\mathrm{{H_{2}O}}$ & 14 & O-H & 1.02 & $[0.1,1.2]$ \\
		$\mathrm{{BeH_{2}}}$ & 14 & Be-H & 1.32 & $[0.1,1.4]$\\
		$\mathrm{{NH_{3}}}$ & 16 & N-H & 1.05 & $[0.1,1.1]$\\
		\hline 
	\end{tabular}
\end{table}

To implement the numerical experiments, we compute electronic integrals,
FCI ground-state energy, and construct the system Hamiltonian using the OpenFermion
package \cite{openfermion} in the STO-3G basis set, simulate quantum
circuits and perform gradient calculations in VQE procedure via the
MindSpore Quantum platform \cite{MindQuantum}, and employ the gradient-based
Broyden--Fletcher--Goldfarb--Shannon (BFGS) algorithm \cite{bfgs}
for parameter optimization. The algorithms (GBEF and ADAPT-VQE) are run on an 16-core 2.10 GHz CPU with 120 GB memory and an RTX 4090 (24GB) GPU.

\begin{table}[b]
	\caption{Parameter reduction  performance benchmarks comparing GBEF (with ablation studies) against
		standard UCCSD, ADAPT-VQE, and SymUCCSD: parameter counts, reduction 
		ratios, and GBEF's truncation thresholds.}\label{tab:bench}
	\centering
	\fontsize{9}{12}\selectfont
	\begin{tabular}{|c|c|c|c|c|c|c|c|c|c|c|c|}
		\hline 
		\multirow{3}{*}{\begin{cellvarwidth}[t]
		\centering
		\!\!\!Mole-\!\!\!\\\!\!\!cule\!\!\!
	\end{cellvarwidth}} & \multicolumn{7}{c|}{Number of Parameters} & \multicolumn{2}{c|}{Reduction$^{**}$} & \multicolumn{2}{c|}{\!Threshold\!}\\
		\cline{2-12}
		& \multirow{2}{*}{\!\!\!UCCSD\!\!\!} & \multicolumn{3}{c|}{\!ADAPT-} & \multirow{2}{*}{\begin{cellvarwidth}[t]
				\centering
				\!\!\!Sym-\!\!\!\\\!\!\!UCCSD\!\!\!
		\end{cellvarwidth}} & \multirow{2}{*}{\!\!\!GBEF\!\!\!} & \multirow{2}{*}{\!\!\!Ablation\!\!\!} & \multirow{2}{*}{\!\!GBEF\!\!} & \multirow{2}{*}{\!\!Ablation\!\!} & \multirow{2}{*}{\!abs\!} & \multirow{2}{*}{mag}\tabularnewline
		\cline{3-5}
		&  & $\epsilon_{1}$$^{*}$ & $\epsilon_{2}$ & $\epsilon_{3}$ &  &  &  &  &  &  & \tabularnewline
		\hline 
		$\mathrm{{H_{4}}}$ & 14 & 5.33(52.38\%) & 6.86 & 7.81 & 8 & 5.52 & 5.33 & 19.53\% & 22.3\% & 0.05 & 0.50\tabularnewline
		\hline 
		$\mathrm{{HF}}$ & 20 & 2.09(90.48\%) & 6.81 & 10.05 & 11 & 4.38 & 2.14 & 35.68\% & 68.58\% & 0.06 & 0.45\tabularnewline
		\hline 
		$\mathrm{{H_{2}O}}$ & 65 & 5.24(19.05\%) & 18.62 & 24.52 & 26 & 10.05 & 8.67 & 46.03\% & 53.44\% & 0.05 & 0.27\tabularnewline
		\hline 
		$\!\mathrm{{BeH_{2}}}\!$ & 90 & 7.33(33.33\%) & 18.90 & 22.71 & 23 & 10.67 & 8.71 & 43.54\% & 53.92\% & 0.04 & 0.30\tabularnewline
		\hline 
		$\mathrm{{NH_{3}}}$ & 135 & 6.67(0\%) & 57.95 & 85.19 & 75 & 44.52 & 32.71 & 23.18\% & 43.36\% & \!0.017\! & 0.40\tabularnewline
		\hline 
	\end{tabular}
	$^{*}$ The ADAPT-$\epsilon_{1}$ algorithm's overly relaxed convergence
	threshold led to out-of-tolerance VQE errors (parentheses show the
	percentage of cases achieving chemical accuracy). In contrast, all
	other methods achieve chemical precision throughout.

	$^{**}$ These percentages are benchmarked against ADAPT-$\epsilon_{2}$'s
	results, which guarantee chemical accuracy throughout and represent
	the lowest parameter count among existing approaches.
\end{table}

Tab.~\ref{tab:bench}  compares the parameter reduction performance of our GBEF algorithm 
with UCCSD, ADAPT-VQE, and SymUCCSD benchmarks, including ablation studies 
and the pruning threshold used in GBEF.
In the ADAPT-VQE algorithms, the convergence threshold $\epsilon_{m}=10^{-m}$ specifies
the termination criterion for the iterative procedure \cite{ADAPT-VQE}.
Generally, a smaller value of $m$ in ADAPT-VQE generates more compact
ansatze with fewer parameters, but risks introducing errors that may
violate the chemical accuracy threshold (defined as a 1.6 mHa deviation
from FCI results) in the final VQE solutions. The GBEF algorithm employs
a dual-threshold parameter truncation strategy in Step 6 that combines
both absolute and relative criteria. First, parameters with gradients
below the absolute threshold ``abs'' are systematically discarded
to eliminate physically insignificant contributions. Subsequently,
the remaining parameters undergo relative importance evaluation, where
the parameter list is truncated at the point where sequential gradients
exhibit a $10^{\text{mag}}$-fold decrease in their norms, thereby
retaining only the most dominant parameters while maintaining computational
efficiency. The thresholds abs and mag can be adjusted according to the gradient list to achieve an optimal balance between circuit depth and solution accuracy. 
This hierarchical approach ensures robust parameter selection
by considering both the absolute physical significance and relative
quantitative importance of each term. 
In the GBEF ablation study, we progressively eliminate parameters with the smallest gradients from the pre-sorted list (obtained in Step 5) until attaining the minimal parameter set that preserves chemical accuracy in VQE results, representing the practical performance limit of the GBEF.

As evidenced in Tab.~\ref{tab:bench}, our GBEF algorithm and its
ablation achieve the best performance, reducing parameter counts by
20-46$\%$ and 22-69$\%$ respectively compared to the ADAPT-$\epsilon_{2}$
algorithm, while maintaining chemical accuracy in all VQE results.
The Ablation results also reveal substantial optimization potential in
our GBEF algorithm, particularly through developing more advanced
truncation protocols for Step 6.

\begin{table}[h]
	\caption{Comparison of final Ansatz circuit depths between the GBEF and ADAPT-VQE
		algorithms. All excitations are implemented using the Jordan-Wigner encoding \cite{Jordan-Wigner}.} \label{tab:depth}
	\centering
	\begin{tabular}{|c|c|c|c|c|c|c|c|}
		\hline 
		\multirow{2}{*}{Molecule} & \multicolumn{3}{c|}{ADAPT-} & \multirow{2}{*}{GBEF} & \multirow{2}{*}{Ablation} & \multicolumn{2}{c|}{Reduction$^*$}\\
		\cline{2-4}\cline{7-8}
		& $\epsilon_{1}$ & $\epsilon_{2}$ & $\epsilon_{3}$ &  &  & GBEF & \multicolumn{1}{c|}{Ablation}\tabularnewline
		\hline 
		\hline 
		$\text{H}_{4}$ & 451.14 & 641.43 & 692.90 & 544.81 & 544.81 & 15.06\% & 15.06\%\tabularnewline
		\hline 
		$\text{HF}$ & 148.71 & 422.14 & 773.57 & 259.24 & 154.10 & 38.59\% & 63.50\%\tabularnewline
		\hline 
		$\text{H}_{2}\text{O}$ & 390.33 & 2299.19 & 3143.67 & 930.29 & 866.62 & 59.54\% & 62.31\%\tabularnewline
		\hline 
		$\text{BeH}_{2}$ & 563.95 & 1780.19 & 2285.76 & 959.14 & 809.33 & 58.04\% & 64.59\%\tabularnewline
		\hline 
		$\text{NH}_{3}$ & 452.43 & 10758.05 & 17041.33 & 8340.62 & 5230.90 & 22.47\% & 51.37\%\tabularnewline
		\hline 
	\end{tabular}\\
	\fontsize{9}{12}\selectfont
	$^*$ The ``Reduction'' results indicate the circuit depth reduction ratios of GBEF and its Ablation relative
	to ADAPT-$\epsilon_{2}$, which represents the state-of-the-art algorithm
	achieving chemical accuracy with the shallowest circuit depth.
\end{table}
Tab.~\ref{tab:depth} compares the final ansatz circuit depths between GBEF (with ablation) and ADAPT-VQE algorithms with different convergence thresholds. We emphasize that the Jordan-Wigner encoding \cite{Jordan-Wigner} is employed here to transform excitations to quantum circuits. While alternative methods could produce shallower circuit, the conclusion still holds.
Tab.~\ref{tab:depth} demonstrates that our GBEF achieves further circuit depth reduction compared to the current best algorithm -- ADAPT-$\epsilon_2$, reaching up to about $60\%$ reduction (with ablation studies showing a $65\%$ upper bound). These results confirm that our GBEF method generates circuits that are more executable on NISQ devices and capable of producing meaningful results, owing to their shallower depth.

Tab.~\ref{tab:runtime} summarizes the runtime performance of GBEF versus ADAPT-VQE algorithms for UCCSD ansatz simplification. The data reveal that as molecular size increases, GBEF demonstrates progressively more significant acceleration over ADAPT-$\epsilon_2$. Notably, GBEF achieves a over $600\times$ speedup  for NH$_3$. Benefiting from its lower computational complexity, the GBEF can efficiently handle UCCSD ansatz simplification for large molecular systems.

\begin{table}[h]
	\centering
	\caption{Comparison of runtime between GBEF and ADAPT-VQE algorithms with different
		convergence thresholds.  All ADAPT-VQE implementations are
		classically simulated due to current quantum hardware limitations.} \label{tab:runtime}
	\begin{tabular}{|c|c|c|c|c|c|}
		\hline 
		\multirow{2}{*}{Molecule} & \multicolumn{3}{c|}{ADAPT-} & \multirow{2}{*}{GBEF} & \multirow{2}{*}{Speedup$^*$ ($\times$)}\tabularnewline
		\cline{2-4}
		& $\epsilon_{1}$ & $\epsilon_{2}$ & $\epsilon_{3}$ &  & \tabularnewline
		\hline 
		\hline 
		$\text{H}_{4}$ & 3.05s & 5.73s & 10.08s & 0.45s & 12.73\tabularnewline
		\hline 
		$\text{HF}$ & 2.66s & 8.38s & 21.75s & 0.91s & 9.21\tabularnewline
		\hline 
		$\text{H}_{2}\text{O}$ & 26.38s & 164.96s & 365.59s & 1.96s & 84.16\tabularnewline
		\hline 
		$\text{BeH}_{2}$ & 42.86s & 145.29s & 314.67s & 1.62s & 89.69\tabularnewline
		\hline 
		$\text{NH}_{3}$ & 471.23s & 6275.84s & 8185.81 & 9.25s & 678.47\tabularnewline
		\hline 
	\end{tabular}\\
	\fontsize{9}{12}\selectfont
	$^*$ The ``Speedup'' values indicate the acceleration
	factor of GBEF relative to ADAPT-$\epsilon_{2}$, as it
	represents the best algorithm achieving chemical accuracy
	with the shallowest circuit depth.
\end{table}

\begin{figure*}[ht!]
	\centering
	\includegraphics[scale=0.64]{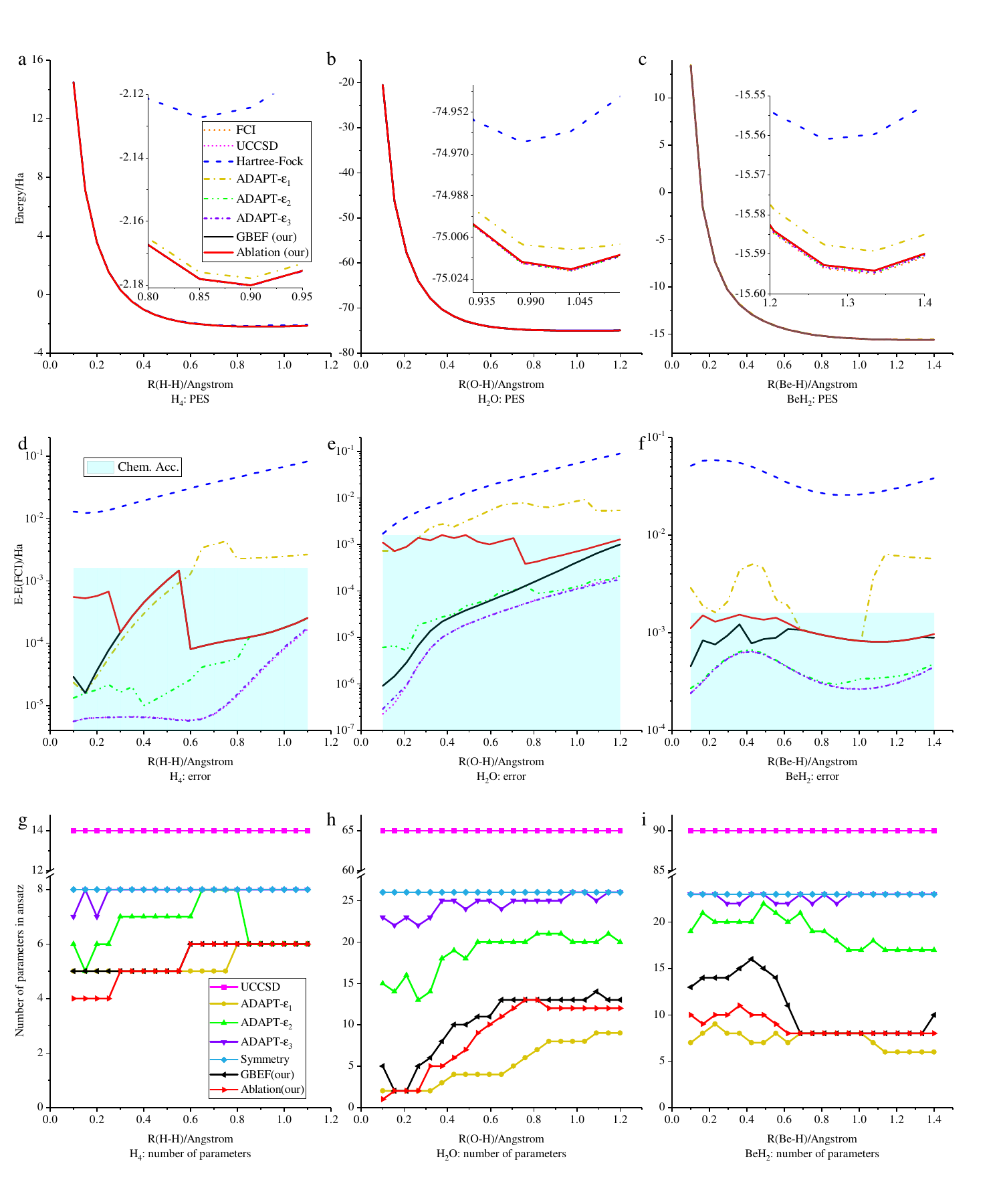}
	\caption{Potential energy surfaces (a-c), solution errors (d-f), and parameter
		count distributions (g-i) versus bond length obtained by different
		algorithms for the H$_4$, H$_2$O, and BeH$_2$
		molecular systems. Shaded blue region represents area within “Chem.
		Acc.” as 1.6 mHa.}\label{fig:main_res}
\end{figure*}

Fig.~\ref{fig:main_res} visually compares the Potential Energy
Surfaces (PES), VQE errors compared with FCI results, and parameter
distributions obtained by different algorithms for three investigated
molecular systems (H$_4$, H$_2$O and BeH$_2$). 
As visualized in Fig.~\ref{fig:main_res}, Hartree-Fock and ADAPT-$\epsilon_{1}$ fail to meet the chemical
accuracy standard, while UCCSD, ADAPT-$\epsilon_{2}$, ADAPT-$\epsilon_{3}$,
GBEF and its Ablation maintain acceptable errors across all bond lengths.
ADAPT-$\epsilon_{3}$ achieves comparable solution accuracy to UCCSD,
but requires substantially more parameters (similar to SymUCCSD).
While ADAPT-$\epsilon_{2}$ currently represents the state-of-the-art method by balancing
accuracy and parameter efficiency, our GBEF and Ablation
achieve superior parameter reduction. A particularly valuable observation
is that within the bond length range of 0.7-1.0 $\mathring{\text{A}}$ for the BeH$_2$ molecule,
both GBEF and ADAPT-$\epsilon_{1}$ take identical parameters, even
though the implementation cost of GBEF is significantly lower than
that of ADAPT-$\epsilon_{1}$.

\section{Conclusion}

This paper presents an efficient method for filtering out non-critical
excitations in the UCCSD ansatz, enabling the construction of shallower
quantum circuits for molecular ground-state energy calculations. We
establish a fundamental connection between the excitations' gradients
at Hartree-Fock state and their correlation energy contributions, and propose
a gradient-based excitation sorting and truncation protocol. By compressing
both the Hamiltonian and Hartree-Fock state into the excitation's subspace,
these gradients can be efficiently  calculated in  parallel on classical
computers. Additionally, we propose grouping the excitations based
on the involved free parameters and utilizing spin-adaptation constraints
to reduce the number of gradient computations. Compared to existing best ADAPT-VQE
algorithm, the GBEF algorithm achieves about $20$-$46\%$ parameter decrease, $15$-$60\%$ circuit depth reduction, and $9$-$678\times$ runtime speedup in the investigated molecules while
maintaining chemical accuracy, demonstrating significantly improved feasibility for near-term experimental applications.

\section{APPENDIX}\label{sec:Appendix}

\textbf{Theorem 1.} \textit{If the Hartree-Fock state $|HF\rangle$ is a good approximation to the exact ground state wave-function $|\Phi_{0}\rangle$, an excitation's contribution to the correlation energy is related to the gradient of the system energy involved only this excitation at the $|HF\rangle$ with respect to the coefficient $\theta=0$.}

\begin{proof}
We refer to the exact ground state of the system in the single- and
double-excitation subspace as $|\Phi_{0}\rangle$, which can be written
as the following form in the configuration interaction expansion:

\begin{equation}
	|\Phi_{0}\rangle=c_{0}|HF\rangle+\sum_{a,r}c_{a}^{r}|\Psi_{a}^{r}\rangle+\sum_{a>b,r>s}c_{ab}^{rs}|\Psi_{ab}^{rs}\rangle,\label{eq:expansion}
\end{equation}
where the singly excited determinants $|\Psi_{a}^{r}\rangle=A_{a}^{r}|HF\rangle$
are obtained by replacing the spin orbital $\chi_{a}$ with $\chi_{r}$ in
the $|HF\rangle$, while the doubly excited determinants
$|\Psi_{ab}^{rs}\rangle=A_{ab}^{rs}|HF\rangle$ involve simultaneous
substitutions of $\chi_{a}\rightarrow\chi_{r}$ and $\chi_{b}\rightarrow\chi_{s}$. 
These determinants obey the orthonormality condition: $\langle \Psi_i | \Psi_j \rangle = \delta_{ij}$.
%

The system Hamiltonian $\mathscr{H}$ satisfies the eigenrelation
with the ground state $|\Phi_{0}\rangle$:
\begin{equation}
	\mathscr{H}|\Phi_{0}\rangle=\mathscr{E}_{0}|\Phi_{0}\rangle,
\end{equation}
with $\mathscr{E}_{0}$ referring to the exact ground energy. By
substracting the term $E_{HF}|\Phi_{0}\rangle$ from both sides, the
above equation transforms to 
\begin{equation}
	(\mathscr{H}-E_{HF})|\Phi_{0}\rangle=(\mathscr{E}_{0}-E_{HF})|\Phi_{0}\rangle=E_{\text{corr}}|\Phi_{0}\rangle,
\end{equation}
where $E_{HF}$ denotes the energy of Hartree-Fock state, i.e., $E_{HF}=\langle HF|\mathscr{H}|HF\rangle$, and  $E_{\text{corr}}$
is the interested electron correlation energy, respectively. If we
multiply this equation by $\langle HF|$, we obtain
\begin{equation}
	\langle HF|(\mathscr{H}-E_{HF})|\Phi_{0}\rangle=E_{\text{corr}}\langle HF|\Phi_{0}\rangle=c_0E_{\text{corr}}.\label{eq:E_corr}
\end{equation}
After, we expand the term $|\Phi_{0}\rangle$ on the left-hand side
using Eq.~(\ref{eq:expansion}), yielding
\begin{equation}
	\begin{aligned}\langle HF|(\mathscr{H}-E_{HF})|\Phi_{0}\rangle & =\langle HF|(\mathscr{H}-E_{HF})\left(c_0|HF\rangle+\sum_{a,r}c_a^r|\Psi_{a}^{r}\rangle+\sum_{a>b,r>s}c_{ab}^{rs}|\Psi_{ab}^{rs}\rangle\right)\\
		& =\sum_{a>b,r>s}c_{ab}^{rs}\langle HF|\mathscr{H}|\Psi_{ab}^{rs}\rangle,
	\end{aligned}
	\label{eq:relation}
\end{equation}
where the Brillouin's theorem ($\langle HF|\mathscr{H}|\Psi_{a}^{r}\rangle=\langle\Psi_{a}^{r}|\mathscr{H}|HF\rangle=0$)  \cite{szabo}
is imposed. Combining Eqs.~(\ref{eq:E_corr}) and (\ref{eq:relation}),
the electron correlation energy $E_{\text{corr}}$ emerges as 
\begin{equation}
	E_{\text{corr}}=\sum_{a>b,r>s}\frac{c_{ab}^{rs}}{c_0}\langle HF|\mathscr{H}|\Psi_{ab}^{rs}\rangle\label{eq:E_corr2}.
\end{equation}
Thus, the electronic correlation energy fundamentally comprises a
weighted summation of interaction terms $\langle HF|\mathscr{H}|\Psi_{ab}^{rs}\rangle$.

For a given excitation $A\in\{\hat{a}_{r}^{\dag}\hat{a}_{a}-\hat{a}_{a}^{\dagger}\hat{a}_{r},\hat{a}_{r}^{\dagger}\hat{a}_{s}^{\dagger}\hat{a}_{a}\hat{a}_{b}-\hat{a}_{a}^{\dagger}\hat{a}_{b}^{\dagger}\hat{a}_{r}\hat{a}_{s}\}$,
the gradient of system energy (or alternatively denoted as the excitation's gradient, for simplicity) with respect to the excitation's coefficient $\theta=0$ is
\begin{equation}
	\frac{\text{d}E}{\text{d}\theta}\Big|_{\theta=0}:=\frac{\text{d}}{\text{d}\theta}\langle HF|\text{e}^{-\theta A}\mathscr{H}\text{e}^{\theta A}|HF\rangle\vert_{\theta=0}=\langle HF|[\mathscr{H},A]|HF\rangle.
\end{equation}
For a double excitation $A_{ab}^{rs}$, this gradient is 
\[
\frac{\text{d}E}{\text{d}\theta}\Big|_{\theta=0}=\langle HF|\mathscr{H}|\Psi_{ab}^{rs}\rangle+\text{h.c.}
\]
Due to the hermiticity of $\mathscr{H}$, for real determinations
$|\Psi_{ab}^{rs}\rangle$ and $|HF\rangle$, $\frac{\text{d}E}{\text{d}\theta}\big|_{\theta=0}=2\langle HF|\mathscr{H}|\Psi_{ab}^{rs}\rangle$.
Combining with Eq.~(\ref{eq:relation}), this result implies that
the gradient of the excitation in $|HF\rangle$ when its associated parameter $\theta=0$
is related to its contribution to the electron correlation
energy $E_{\text{corr}}$. If the $|HF\rangle$ is a good approximation to the $|\Phi_{0}\rangle$, its coefficient $c_0$ will be larger than any of others in the configuration interaction expansion Eq.~(\ref{eq:expansion}), leading to the weights in Eq.~(\ref{eq:E_corr2}) $\big|\frac{c_{ab}^{rs}}{c_0}\big|\le 1$. 
If $|\langle HF|\mathscr{H}|\Psi_{ab}^{rs}\rangle|\approx 0$, the corresponding contribution $\frac{c_{ab}^{rs}}{c_0}\langle HF|\mathscr{H}|\Psi_{ab}^{rs}\rangle$ in Eq.~(\ref{eq:E_corr2}) becomes negligible. This implies that the gradient of a double excitation when its parameter $\theta=0$ reflects its contribution to the correlation energy.

Analogously, the gradient of a single excitation $A_{a}^{r}$
at $|HF\rangle$ when $\theta=0$ is $\frac{\text{d}E}{\text{d}\theta}\big|_{\theta=0}=2\langle HF|\mathscr{H}|\Psi_{a}^{r}\rangle=0$, because of the Brillouin's theorem.
This result is consistent with the empirical observation that the
inclusion of single excitations often contributes negligibly
to ground-state energy calculations. 
\end{proof}

In some more refined computational scenarios, the contributions
of single excitations have to be taken into account. Although
single excited states do not directly interact with the Hartree-Fock
state, they indirectly influence the system's energy by modifying
the coefficients of double excited states, since $\langle\Psi_{a}^{r}|\mathscr{H}|\Psi_{bc}^{st}\rangle\ne0$
when $a\in\{b,c\}$. 

\section*{Data and code availability \label{sec:data-and-code}}
The code and data used or presented
in this paper are openly available on Gitee at \url{https://gitee.com/mindspore/mindquantum/tree/research/paper_with_code}.

\ack
The authors thank Junyuan Zhou, Keli Zheng, Dingchao Gao, Riling Li for their valuable discussion.
This work is sponsored by Innovation Program for Quantum Science and Technology under Grant No. 2024ZD0300502,  Beijing Nova Program Grant No. 20220484128 and 20240484652,
and CPS-Yangtze Delta Region Industrial Innovation
Center of Quantum and Information Technology-MindSpore Quantum Open Fund.

\section*{References}
\bibliographystyle{unsrt}
\bibliography{References_library}

\section*{Author contributions}
\noindent
%
RH He, and SG Ying established the key idea in this paper. RH He and QZ Chai did the numerical experiments. RH He, QZ Chai and X Hong wrote the first version of the draft paper. All the authors contributed to the preparation of this paper.\\
\section*{Competing interests}
%
%
The authors declare no competing interests.\\
\section*{Correspondence}
%
\textbf{Correspondence} and requests for materials should be addressed to RH He or SG Ying.

\end{document}